\documentclass{iopart}
\usepackage{iopams}
\usepackage{amssymb}
\usepackage{bm}
\usepackage{graphicx}
\begin{document}

\title{Local and nonlocal observables in quantum optics}
\author{Iwo Bialynicki-Birula}
\address{Center for Theoretical Physics, Polish Academy of Sciences,\\Al. Lotnik\'ow 32/46, 02-668 Warsaw, Poland}
\ead{birula@cft.edu.pl}

\begin{abstract}
It is pointed out that there exists an unambiguous definition of locality that enables one to distinguish local and nonlocal quantities. Observables of both types coexist in quantum optics but one must be very careful when attempting to measure them. A nonlocal observable which formally depends on the spatial position $\bi r$ cannot be {\em locally} measured without disturbing the measurements of this observable at all other positions.
\end{abstract}
\noindent{\em Keywords\/}: local observables, angular momentum of light, quantum mechanics of photons
\pacs{42.50.-p,42.50.Tx, 03.50.De}
\submitto{New Journal of Physics}

\section{Introduction}

In a recent publication Bliokh, Dressel, and Nori \cite{bdn} addressed the problem of local conservation laws in quantum optics. They claim to have constructed ``{\em local} spin and orbital angular momentum densities and fluxes that satisfy the proper continuity equations'' and they devote a lot of attention to this problem (the word {\em local} appears 32 times in the paper). Since optical experiments play a dominant role in the present day studies of the foundations of quantum mechanics, it is of great importance to understand at the fundamental level all the issues related to locality vs. nonlocality. These problems acquire also an additional significance in the context of Bell's theorem where the notion of locality appears in a profound way.

One cannot overestimate the significance of locality in modern physics. In particular, as was shown for the first time by Pauli \cite{pauli}, the assumption of locality implies one of the most fundamental laws of physics: the connection between spin and statistics. Particles with half-integer spin are fermions and those with integer spin are bosons. Local properties of quantum fields are a trademark of relativistic field theory \cite{rh}. They led in the past to two vast areas of research in high energy physics: the dispersion relations for scattering amplitudes and the current algebras. Quantum optics is part of quantum electrodynamics and it inherits from QED its fundamental features; one of them is the notion of locality.

In this paper I make a systematic use of the fact that there is no arbitrariness in identifying local and nonlocal quantities because there is a very well established, unambiguous {\em criterion of locality}. Every relativistic field theory must obey this criterion. Bjorken and Drell in their classic textbook \cite{bd} give the following succinct definition of locality: ``According to the condition of microscopic causality, local densities ${\mathcal O}(x)$ of observable operator quantities
\begin{eqnarray}\label{bd}
{\mathcal O}\equiv\int\!d^3r\,{\mathcal O}({\bi x},t)
\end{eqnarray}
do not interfere and therefore are required to commute for space-like separations; that is
\begin{eqnarray}\label{comm}
\left[{\mathcal O}(x),{\mathcal O}(y)\right]\equiv 0\quad {\rm for}\;(x-y)^2<0\;''
\end{eqnarray}
In other words the measurement of the observable ${\mathcal O}$ at the point $\bi r$ has no influence on the measurement performed at the point $\bi r'$; the two observables are compatible.

Unfortunately, as is shown in detail in the next Section, the quantities called ``{\em local} spin and orbital angular momentum densities'' in \cite{bdn} fail completely the test of locality. I want to emphasize that the mere splitting of the total angular momentum into its orbital and spin parts introduced in \cite{bdn} does not present any problems. This has been done long time ago for the electromagnetic field by Darwin \cite{dar} and rephrased recently in the framework of the quantum mechanics of photons \cite{bb0}. It is only the claim that such a splitting leads to {\em local densities} that is invalid.

\section{Locality and relativistic invariance}

One may formulate the criterion of locality even in classical electrodynamics in terms of Poisson brackets, but the use of commutators seems to be more appropriate since they are directly related to measurements of quantum observables. Angular momentum in field theory is constructed from the components of the energy-stress tensor. Thus, it is natural to start the analysis from the description of the local properties of this tensor. The fundamental features of all relativistic field theories are the commutator relations satisfied by the components of the energy-stress tensor which guarantee relativistic invariance of the theory \cite{dir,js,bd1}. For the electromagnetic field these relations have the following form \cite{ibb1,dm,bb}:
\numparts
\begin{eqnarray}\label{esrel}
\fl\qquad\left[T^{00}({\bi r},t)\,T^{00}({\bi r}',t)\right]=-\rmi\hbar \left(T^{0k}({\bi r},t)+T^{0k}({\bi r}',t)\right)\partial_k\delta^{(3)}({\bi r}-{\bi r'}),\\
\fl\qquad\left[T^{00}({\bi r},t)\,T^{0k}({\bi r}',t)\right]=-\rmi\hbar \left(T^{ki}({\bi r},t)+T^{00}({\bi r}',t)\delta^{ki}\right)\partial_i\delta^{(3)}({\bi r}-{\bi r'}),\\
\fl\qquad\left[T^{0k}({\bi r},t)\,T^{0l}({\bi r}',t)\right]=-\rmi\hbar \left(T^{0l}({\bi r},t)\partial_k+T^{0k}({\bi r}',t)\partial_l\right)\delta^{(3)}({\bi r}-{\bi r'}).
\end{eqnarray}
\endnumparts
The presence of the three-dimensional delta functions on the right hand side in these equations is a clear indication of locality. Energy density $T^{00}({\bi r},t)$ and momentum density $T^{0k}({\bi r},t)$ are clearly {\em local variables}. In contrast, this property is not shared by the components of the canonical energy-stress tensor. For example, the canonical momentum density $E_k\nabla A_k^\perp$ employed to construct the orbital angular momentum in \cite{bdn} is a nonlocal object. The nonlocality is due to the presence of the transverse vector potential. The commutation relations involving this vector contain the nonlocal term,
\begin{eqnarray}\label{cancomm}
\left[A_i^\perp({\bi r}),D_j({\bi r}')\right]=
-\rmi\hbar\delta_{ij}\delta^{(3)}({\bi r}-{\bi r}')
-\frac{\rmi\hbar}{4\pi}\partial_i\partial_j\frac{1}{|{\bi r}-{\bi r'}|}.
\end{eqnarray}
I use ${\bi D}$ rather than ${\bi E}$ because ${\bi B}$ and ${\bi D}$ are the true canonical variables; their commutation relations do not contain any material constants and they have the following universal form:
\begin{eqnarray}\label{cancomm0}
\left[B_i({\bi r}),D_j({\bi r}')\right]
=\rmi\hbar\epsilon_{ijk}\partial_k\delta^{(3)}({\bi r}-{\bi r}').
\end{eqnarray}
In Maxwell theory the distinction between  ${\bi D}$ and ${\bi E}$ is less important but it becomes crucial in a nonlinear theory of electromagnetism \cite{bi}.

The presence of the nonlocal term in (\ref{cancomm}) is easily explained; the transverse part of the potential is a {\em nonlocal} function of the {\em local} field ${\bi B}$,
\begin{eqnarray}\label{vpot}
{\bi A}^\perp({\bi r})={\bi\nabla}\times\int\!\frac{d^3r'}{4\pi}\frac{{\bi B}({\bi r'})}{|{\bi r}-{\bi r'}|}.
\end{eqnarray}
Therefore, local commutation relations (\ref{cancomm0}) between ${\bi B}$ and ${\bi D}$ result in nonlocal commutation relations between ${\bi A}^\perp$ and ${\bi D}$.

\section{Important nonlocal quantities}

Of course, it is not forbidden to use nonlocal quantities in quantum optics; they should be just recognized as such. Some of them might be useful and some of them are even of fundamental significance. The total number of photons $N$ and the total helicity $\Lambda$ are the most important examples,
\begin{eqnarray}\label{totn}
\fl\qquad N=\frac{1}{4\pi^2\hbar c}\int\!\!d^3r\int\!\!d^3r'\!\frac{1}{\vert{\bi r}-{\bi r}'\vert^2}\left[\frac{1}{\epsilon}{\bi D}({\bi r})\!\cdot\!{\bi D}({\bi r}')+\frac{1}{\mu} {\bi B}({\bi r})\!\cdot\!{\bi B}({\bi r}')\right].
\end{eqnarray}
\begin{eqnarray}\label{totl}
\fl\qquad \Lambda=\frac{1}{4\pi^2\hbar c}\int\!\!d^3r\int\!\!d^3r'\!\frac{1}{\vert{\bi r}-{\bi r}'\vert}\left[\frac{1}{\epsilon}{\bi D}({\bi r})\!\cdot\!{\bi\nabla}\times{\bi D}({\bi r}')+\frac{1}{\mu} {\bi B}({\bi r})\!\cdot\!{\bi\nabla}\times{\bi B}({\bi r}')\right].
\end{eqnarray}
The total number of photons, introduced in \cite{zeld}, plays an important role in the statistical theory of the electromagnetic field formulated in terms of the Wigner function \cite{ibb2}. As was to be expected, the total number of photons is an absolute scalar. It is invariant not only under all Poincar\'e transformations but also under conformal transformations \cite{lg}. The total helicity is also an invariant and it is the generator of duality transformation \cite{dt}.

The total number of photons may be viewed as a space integral of the density of photons $n({\bi r},t)$,
\begin{eqnarray}\label{nphot}
\fl\qquad n({\bi r},t)=\frac{1}{4\pi^2\hbar c}\int\!\!d^3r'\!\frac{1}{\vert{\bi r}-{\bi r}'\vert^2}\left[ \frac{1}{\epsilon}{\bi D}({\bi r})\!\cdot\!{\bi D}({\bi r}')+\frac{1}{\mu} {\bi B}({\bi r})\!\cdot\!{\bi B}({\bi r}')\right].
\end{eqnarray}
This nonlocal density of photons satisfies a continuity equation $\partial_t n+\nabla\cdot{\bi j}=0$ with the nonlocal photon current ${\bi j}({\bi r},t)$,
\begin{eqnarray}\label{curr}
\fl\qquad{\bi j}({\bi r},t)=\frac{c}{4\pi^2\hbar}\int\!\!d^3r'\!\frac{1}{\vert{\bi r}-{\bi r}'\vert^2}\left[{\bi D}({\bi r})\times{\bi B}({\bi r}')
-{\bi B}({\bi r})\times{\bi D}({\bi r}')\right].
\end{eqnarray}
These formula show clearly that $n({\bi r},t)$ and ${\bi j}({\bi r},t)$ are nonlocal quantities. Since they depend on the electromagnetic fields in the whole space, their measurements in spatially separated regions are not independent. Analogous nonlocal structures, namely the spin and angular momentum densities, were introduced in \cite{bdn}. These objects, like $n({\bi r},t)$, are indeed the functions of position ${\bi r}$ but that fact does not make them local.

\section{The splitting of the angular momentum into two parts}

The total angular momentum of the electromagnetic field is an integral of the local density of angular momentum. This density is the spatial part of the relativistic tensor $M^{\mu\nu\lambda}$,
\begin{eqnarray}\label{angm}
M^{\mu\nu\lambda}=x^\mu T^{\nu\lambda}-x^\nu T^{\mu\lambda}.
\end{eqnarray}
Owing to the symmetry of the energy-stress tensor, $M^{\mu\nu\lambda}$ satisfies the {\em local} continuity equation (cf., for example, \cite{bb0} p. 92),
\begin{eqnarray}\label{relcont}
\partial_\lambda M^{\mu\nu\lambda}=0.
\end{eqnarray}
In particular, the spatial part of $M^{\mu\nu\lambda}$ satisfies the continuity equation of the form:
\begin{eqnarray}\label{ceam}
\partial_tM^{ij}+\partial_k(x^iT^{jk}-x^jT^{ik})=0,
\end{eqnarray}
where $T^{ij}$ is the Maxwell stress tensor. This continuity equation is local but when one splits the total angular momentum into its orbital part and the intrinsic part, locality cannot be preserved. This is clearly seen from the formulas for the total angular momentum ${\bi J}$.
\begin{eqnarray}\label{totam}
J_k=\frac{1}{2}\epsilon_{kij}\int\!d^3r\,M^{ij}({\bi r}).
\end{eqnarray}
Even though ${\bi J}$ is an integral of a local density of angular momentum, after the splitting, ${\bi J}={\bi J}_O+{\bi J}_S$, the locality is lost. Indeed, following \cite{dar}, one obtains:
\numparts
\begin{eqnarray}\label{darwin2}
{\bi J}_O=\int\!d^3rD_i({\bi r})({\bi r}\times{\bi\nabla})A_i^\perp({\bi r}),\\
{\bi J}_S=\int\!d^3r{\bi D}({\bi r})\times{\bi A}^\perp({\bi r}).
\end{eqnarray}
\endnumparts
The integrands (called orbital and spin densities in \cite{bdn}) when expressed in terms of potentials look local, but they contain a hidden nonlocality. Namely, when taken at two different points, they do not commute. I will not give the explicit formulas for the commutators here because they are fairly complicated but it suffices to observe that according to (\ref{cancomm}) they always contain the nonlocal terms $\partial_i\partial_j(1/|{\bi r}-{\bi r'}|)$. The nonlocality becomes even more explicit when the formulas (\ref{darwin2}) are rewritten in terms of local fields,
\numparts
\begin{eqnarray}\label{darwin3}
{\bi J}_O=\int\!d^3r\int\!d^3r'D_i({\bi r})\left({\bi r}\times{\bi\nabla}\right)\frac{\left({\bi\nabla}'\times{\bi B}({\bi r}')\right)_i}{4\pi|{\bi r}-{\bi r'}|},\\
{\bi J}_S=\int\!d^3r\int\!d^3r'\frac{{\bi D}({\bi r})\times\left({\bi\nabla}'\times{\bi B}({\bi r}')\right)}{4\pi|{\bi r}-{\bi r'}|}.
\end{eqnarray}
\endnumparts
Therefore, the separate measurement of any of these two densities in the vicinity of $\bi r$ influences the measurements everywhere. This fact invalidates the claims made in \cite{bdn} that ``this separation produces a meaningful local description of the spin and orbital AM densities''.

\section{Origins of nonlocality}

All nonlocal quantities in the theory of electromagnetism have a common cause. They originate from the use the Fourier transformation. This is already seen in the definition of the transverse part ${\bi A}^\perp$ of the vector potential. In terms of the Fourier transform, the transverse part of ${\tilde{\bi A}}(\bi k)$ is obtained by an algebraic operation:
\begin{eqnarray}\label{alg}
{\tilde{\bi A}}^\perp(\bi k)=-\frac{{\bi k}\times{\bi k}\times{\tilde{\bi A}}(\bi k)}{{\bi k}^2}.
\end{eqnarray}
While multiplication in this formula by ${\bi k}$ simply produces derivatives in coordinate space, the division by ${\bi k}^2$ results in a nonlocal expression (\ref{vpot}). The nonlocality in the formula (\ref{totn}) for the photon number has the same origin. The energy density is a local quantity but to obtain the photon number one has to divide by $\hbar ck$ and this can be done only in the Fourier space. The same explanation holds for the orbital angular momentum. Its separation from the total angular momentum requires taking the part that is orthogonal to ${\bi k}$; the part parallel to ${\bi k}$ is the intrinsic (helicity) part as shown in \cite{bb0}. These operations are again algebraic in the Fourier space but are nonlocal in the coordinate space.

\section{Conclusions}

The meaning of locality in a relativistic field theory has a precise meaning. This is especially true in the quantum theory of the Maxwell field in the vacuum because this theory does not have a nonrelativistic limit and all requirements of relativity strictly apply. A failure to distinguish between local and nonlocal observables may lead to a misinterpretation of observations in quantum optics.

\section*{Acknowledgments}

I would like to thank Zofia Bialynicka-Birula for reading the manuscript and critical remarks. This research was financed by the Polish National Science Center Grant No. 2012/07/B/ST1/03347.


\section*{References}
\begin{thebibliography}{99}
\bibitem{bdn} Bliokh K Y, Dressel J and Nori F 2014 Conservation of the spin and orbital angular momenta in electromagnetism {\em arXiv:} 1404.5486
\bibitem{pauli} Pauli W 1940 The connection between spin and statistics {\em Phys. Rev.} {\bf 58} 716
\bibitem{rh} Haag R 1993 {\em Local Quantum Physis} (Berlin: Springer)
\bibitem{bd} Bjorken J D and Drell S D 1965 {\em Relativistic Quantum Fields} (New York: McGraw-Hill) p. 170
\bibitem{dar} Darwin C G Notes on the theory of radiation 1932 {\em Proc. Roy. Soc. A London} {\bf 136} 36
\bibitem{bb0} Bialynicki-Birula I and Bialynicka-Birula Z 2011 Canonical separation of angular momentum of light into its orbital and spin parts {\em J. Opt.} {\bf 13} 064014
\bibitem{dir} Dirac P A M The condition for a quantum field theory to be relativistic {\em Rev. Mod. Phys.} {\bf 34} 592
\bibitem{js} Schwinger J Energy momentum density in field theory 1963 {\em Phys. Rev.} {\bf 130} 800
\bibitem{bd1} Boulware D G and Deser S 1967 Stress-tensor commutators and Schwinger terms {\em J. Math. Phys.} {\bf 8} 1468
\bibitem{ibb1} Bialynicki-Birula I 1965 Commutation relations for energy-momentum tensor {\em Nuovo Cimento} {\bf 35} 697
\bibitem{dm} Deser S and Morrison K L 1970 Stress-tensor commutators in nonlinear electrodynamics {\em J. Math. Phys.} {\bf 11} 596
\bibitem{bb} Bialynicki-Birula I and Bialynicka-Birula Z 1975 {\em Quantum Electrodynamics} (Oxford: Pergamon) p. 100
\bibitem{bi} Born M and Infeld L 1934 On the quantization of the new field equations I {\em Proc. Roy. Soc. A London} {\bf 147} 522
\bibitem{bb1} Bialynicki-Birula I and Bialynicka-Birula Z 2009 Why photons cannot be sharply localized {\em Phys. Rev. A} {\bf 79} 2383 032112
\bibitem{zeld} Zeldovich Ya B 1965 Number of quanta as an invariant of the classical electromagnetic field {\em Dokl. Acad. Sci. USSR} {\bf 163} 1359 (in Russian)
\bibitem{ibb2} Bialynicki-Birula I 2000 The Wigner functional of the electromagnetic field {\em Opt. Comm.} {\bf 179} 237
\bibitem{lg} Gross L 1964 Norm invariance of mass-zero equations under the conformal group {\em J. Math. Phys.} {\bf 5} 687
\bibitem{dt} Deser S and Teitelboim C 1976 Duality transformations of Abelian and non-Abelian gauge fields {\em Phys. Rev. D} 1592
\end{thebibliography}
\end{document}